\documentclass[lettersize,journal]{IEEEtran}
\usepackage{amsmath,amsfonts}
\usepackage{algorithmic}
\usepackage{algorithm}
\usepackage{array}
\usepackage[caption=false,font=normalsize,labelfont=sf,textfont=sf]{subfig}
\usepackage{textcomp}
\usepackage{stfloats}
\usepackage{url}
\usepackage{verbatim}
\usepackage{graphicx}
\usepackage{cite}

\usepackage{amsmath,amssymb,amsfonts}
\usepackage[hidelinks]{hyperref}
\usepackage{bm}
\usepackage{tabularx}
\usepackage{multirow}

\usepackage[margin=0.55in]{geometry}

\begin{document}
\begin{sloppypar}

\title{Physics-Informed Recurrent Network for State-Space Modeling of Gas Pipeline Networks}

\author{Siyuan Wang, \IEEEmembership{Member, IEEE},
Wenchuan Wu, \IEEEmembership{Fellow, IEEE},
Chenhui Lin, \IEEEmembership{Member, IEEE},
Qi Wang, \IEEEmembership{Member, IEEE},
Shuwei Xu, \IEEEmembership{Member, IEEE},
Binbin Chen
\thanks{This work was supported by the Science-Technique Plan Project of Jiangsu Province (No. SBE2022110130) and the Beijing Natural Science Foundation (L243003). Corresponding author: Wenchuan Wu (e-mail: wuwench@mail.tsinghua.edu.cn)}
\thanks{S.Wang, W. Wu, C. Lin, and S. Xu are with State Key Laboratory of Power Systems, Department of Electrical Engineering, Tsinghua University, Beijing 100084, China.}
\thanks{Q. Wang is with Hong Kong Polytechnic University, Hong Kong Special Administrative Region, China}
\thanks{B. Chen is with ByteDance Ltd., Beijing 100083, China}
}

\markboth{xxxx,~Vol.~xx, No.~x, xxx~xxxx}%
{Shell \MakeLowercase{S. Wang \textit{et al.}}: Physics-Informed Recurrent Network for State-Space Modeling of Gas Pipeline Networks}


\maketitle

\begin{abstract}
As a part of the integrated energy system (IES), gas pipeline networks can provide additional flexibility to power systems through coordinated optimal dispatch. 
An accurate pipeline network model is critical for the optimal operation and control of IESs. However, inaccuracies or unavailability of accurate pipeline parameters often introduce errors in the state-space models of such networks. 
This paper proposes a physics-informed recurrent network (PIRN) to identify the state-space model of gas pipelines. 
It fuses sparse measurement data with fluid-dynamic behavior expressed by partial differential equations.
By embedding the physical state-space model within the recurrent network, parameter identification becomes an end-to-end PIRN training task. 
The model can be realized in PyTorch through modifications to a standard RNN backbone.
Case studies demonstrate that our proposed PIRN can accurately estimate gas pipeline models from sparse terminal node measurements, providing robust performance and significantly higher parameter efficiency. 
Furthermore, the identified state-space model of the pipeline network can be seamlessly integrated into optimization frameworks.
\end{abstract}


\section{Introduction}
\label{sec-Introduction}

\subsection{Background and Motivation} 

Advancements in integrated energy systems have focused on optimizing energy conversion to enhance efficiency and lower carbon emissions.
% \cite{liangCoordinatedSchedulingElectricity2023},  
% \cite{cleggIntegratedElectricalGas2016} 
Gas-fired units and power-to-gas devices are key examples of these technologies and have seen widespread deployment.
% \cite{liOptimalOperationStrategy2018} 
To ensure economically optimal operation, the coordinated dispatch of gas and power systems is essential. 
% \cite{zhangHourlyElectricityDemand2016}
However, inaccuracies in the mathematical modeling of gas pipeline networks can render the results of gas system optimal dispatch calculations infeasible.

In the context of optimal dispatch, it is critical to account for the dynamic behavior of gas flow \cite{zhangDynamicEnergyFlow2022}, 
which is governed by the partial differential equations (PDEs) of fluid dynamics.
% \cite{chenGeneralizedPhasorModeling2021} 
The parameters that define this dynamic process are directly influenced by the gas pipeline network's characteristics. A promising alternative to deriving these parameters analytically is the use of data-driven approaches  \cite{lanDatadrivenStateEstimation2022}. 

Recurrent Neural Networks (RNNs) are a commonly used type of artificial neural network, particularly well-suited for processing sequential and time-dependent data \cite{nakipRenewableEnergyManagement2023}. 
Their architecture includes directed cycles, which enable RNNs to maintain a hidden state, effectively capturing and retaining information from previous inputs. 
This memory capability makes RNNs highly appropriate for simulating dynamic, state-dependent systems, such as gas pipeline networks.
However, the inherent black-box nature of traditional neural networks presents challenges in terms of interpretability, making it difficult to integrate them seamlessly into optimization problems, such as those encountered in optimal dispatch.

In recent years, physics-informed neural networks (PINNs) \cite{raissiPhysicsinformedNeuralNetworks2019} have garnered significant attention within the research community. 
A PINN can seamlessly integrate domain knowledge, particularly physics-based equations, into the classic neural network framework, making it a powerful tool for the simulation, parameter identification, and control of complex physical systems \cite{karniadakisPhysicsinformedMachineLearning2021}. 
This physics knowledge can inform the design of loss functions, initialization of model parameters, and the architecture of the neural network \cite{huangApplicationsPhysicsInformedNeural2023}. 
The inspiration from PINNs provides possible solutions to leverage both the data-driven method and the physics model to identify the parameters of a gas pipeline network.

This paper introduces a physics-informed recurrent network (PIRN) for learning the state-space model of gas pipeline networks. 
By embedding the physical state-space dynamics into each recurrent unit, PIRN unites gray-box interpretability with the capacity to infer full internal states from sparse terminal measurements. 
Its streamlined architecture and transparent structure further enable the identified model to integrate seamlessly into downstream optimization tasks.

\subsection{Literature Review and Research Gap} 

Previous studies have explored the parameter estimation of gas pipeline systems, which is critical for their efficient and safe operation. 
These methods can be broadly classified into several categories: optimization-based methods, Kalman filtering techniques, neural network models, and other miscellaneous approaches. 

Optimization-based approaches aim to identify parameters by minimizing an objective function that quantifies error. 
For instance, a reduced-order model was developed in \cite{sundarDynamicStateParameter2019},
% and \cite{sundarStateParameterEstimation2019}, 
where nonlinear programming and least squares objective function were employed to estimate both the dynamic states and parameters of gas pipeline networks. 
Similarly, maximum likelihood and least squares methods were utilized to identify parameters in \cite{chenIdentificationCharacteristicParameters2022}.
A Bayesian approach to parameter estimation, specifically targeting the friction coefficient in gas networks, was proposed in \cite{hajianBayesianApproachParameter2019}. 
Additionally, \cite{sukharevIdentificationModelFlow2021} employed a lumped-parameter model under both stationary and non-stationary gas flows to estimate pipeline parameters. 
The work in \cite{cuiDatadrivenComputationNatural2020} presented an optimization problem to determine the steady-state of gas networks, where a neural network was used to initialize the optimization process and expedite its solution.

Neural networks have also been explored extensively for parameter estimation, capitalizing on their ability to handle dynamic processes in gas systems. 
A gray-box neural network model was presented in \cite{cenGrayboxNeuralNetworkbased2013} to identify faults within gas systems. 
Long Short-Term Memory (LSTM) models, known for their proficiency in sequence-based data, were applied to predict pipeline shutdown pressures in \cite{zhengDeeppipeTheoryguidedLSTM2021}. 
In addition, a shortcut Elman network was proposed in \cite{zhouDynamicSimulationNatural2022} for dynamic simulations of gas networks.

Kalman filtering is another effective method, particularly for dealing with noisy measurements in parameter estimation. 
For example, \cite{chenDynamicStateEstimation2022}
% \cite{chenRobustDynamicState2022, chenDynamicStateEstimation2022} 
utilized the Kalman filter to estimate the dynamic states of natural gas pipelines. 
Furthermore, a robust Kalman filter-based approach, designed to mitigate the impact of poor-quality data, was introduced in \cite{chenRobustKalmanFilterBased2021}. 
The work in \cite{suGraphFrequencyDomainKalman2024} proposed a graph-frequency domain Kalman filter that reduces the influence of outliers in the data.

Several additional techniques have been proposed for the identification of gas network models and state estimation.
A data-driven approach, developed in \cite{huangDataDrivenStateEstimation2023}, involves solving a weighted low-rank approximation problem to estimate the gas state in the presence of measurement noise. 
Another method, based on Fourier transformation, was presented in \cite{yinEnergyCircuitTheory2020}. This approach transforms the partial differential equations governing gas dynamics from the time domain to the frequency domain, yielding algebraic equations that can be used for state estimation.

In summary, much of the existing research concentrates on simulating the dynamic behavior of gas networks, often integrating physical information to improve model performance \cite{yinHighaccuracyOnlineTransient2023}.
% \cite{suHybridPhysicalData2021, yinHighaccuracyOnlineTransient2023}.
In contrast, relatively little attention has been devoted to parameter identification in gas pipeline networks, a task that fundamentally involves reversing the simulation process.
Moreover, current data-driven approaches to parameter estimation typically emphasize steady-state conditions, offering limited treatment of dynamic state transitions, which are often oversimplified or entirely neglected.
As a result, these methods face significant limitations when applied to real-world gas system operations.
For example, the widely used quasi-dynamic models, which rely on terminal node states to represent the entire pipeline, can introduce significant inaccuracies in optimal dispatch problems \cite{yangEffectNaturalGas2018}. 
Additionally, traditional neural network models face challenges in safety-critical applications due to their poor interpretability. 
Thus, developing interpretable models capable of accurately capturing the physical dynamics of gas pipeline networks is critical for improving operational efficiency and safety.

\subsection{Methods and Contributions}

To identify the state-space model of gas pipeline networks and integrate the results into optimization problems, this paper introduces a PIRN model. 
Unlike classic RNNs that rely on artificial neural perceptron units, our model incorporates a physical state-space representation of the gas pipeline network. 
This transforms the black-box perceptron units in classic RNNs into a physically interpretable gray-box model, where the parameters correspond directly to those in the dynamic equation matrices, providing a higher degree of interpretability. 
In this framework, the task of identifying pipeline parameters is reformulated as a training process for the PIRN. 
Leveraging the memory capability of RNN architecture, our method requires only sparse measurements from the terminal nodes of the gas pipeline, while the internal states along the pipeline are estimated and stored in the hidden layers of the network.
Embedding prior physical knowledge prevents the network from converging to physically unrealistic local optima, ensuring both stability and physical interpretability. Consequently, the identified state-space model can be seamlessly integrated into any optimal dispatch problem for gas networks.

To the best knowledge of the authors, the contributions of this work are as follows:

(1) We introduce a novel physics-informed recurrent network (PIRN) architecture for learning the state-space dynamics of gas pipeline networks. Rather than using generic perceptron units, our approach replaces them with physical state-space functions derived from the governing gas-flow equations, casting model identification as a structured, physics-grounded learning task.

(2) The proposed PIRN requires only sparse measurements at pipeline terminal nodes. Internal gas states are inferred and encoded within the network's hidden layers, enabling full-state reconstruction from highly sparse observations.

(3) Our model integrates seamlessly with the PyTorch framework, requiring only minor adaptations to classic RNN structures. Compared to the classic RNNs, it achieves superior robustness, enhanced interpretability, and significantly improved parameter efficiency.

The remainder of this paper is organized as follows. Section \ref{sec-ProblemFormation} provides an overview of the problem formulation and methodology. The detailed state-space model identification method is presented in Section \ref{sce-Methodology}. Numerical tests are performed in Section \ref{sec-NumericalTests} and conclusions are made in Section \ref{sec-Consclusion}.

%% -----------------------------
%% Problem Formation and Methodology Overview
%% -----------------------------

\section{Problem Formation}
\label{sec-ProblemFormation}

\begin{figure}[h] 
	\centering
	\includegraphics[width=0.35\textwidth]{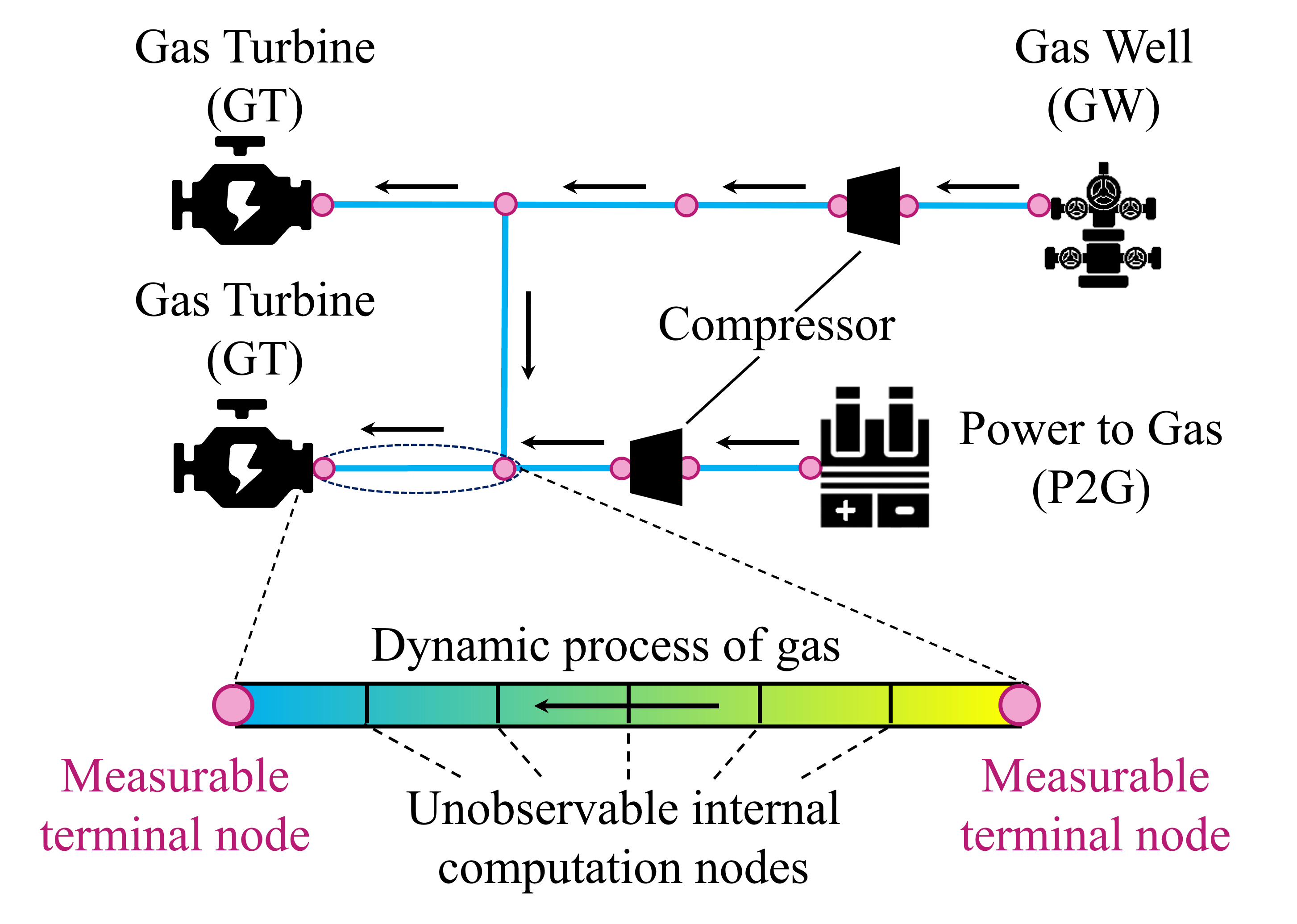}
	\caption{Schematic of the gas pipeline network and its measurement configuration} 
	\label{fig-Schematic}
\end{figure}

The diagram of the gas pipeline network and its measurement configuration are illustrated in Fig.~\ref{fig-Schematic}. 
Natural gas originates from gas wells (GWs) and power-to-gas devices (P2Gs). 
Pressurized by compressors, natural gas is transported to end-users, such as gas turbines (GTs), via the pipeline network. 
The transportation of gas through the pipelines is a relatively slow dynamic process \cite{zhouEquivalentModelGas2017}. 
The gas pressure and mass flow rate at a specific time and position are two physical quantities that characterize the dynamic states of gas within the pipeline network. 
These states are governed by the following fluid partial differential equations:

\begin{equation} \label{eq-PDE1}
	\frac{\partial \rho}{\partial t}+\frac{\partial \rho v}{\partial x}=0
\end{equation}

\begin{equation} \label{eq-PDE2}
	\frac{\partial \rho v}{\partial t}+\frac{\partial \rho v^2}{\partial x}+\frac{\partial \pi}{\partial x}+\frac{\lambda \rho v^2}{2 D}+\rho g \sin \alpha=0
\end{equation}

\noindent where $\rho$, $v$ and $\pi$ denote the density, velocity and pressure of gas, respectively; $v$ is supposed to be greater than 0; $D$, $\lambda$, and $\alpha$ denote the cross-sectional diameter, friction coefficient and inclination of the pipeline, respectively.
 
To calculate the dynamic states of a gas pipeline network for model predictive control, essential parameters include cross-sectional diameter $D$, friction coefficient $\lambda$, and pipeline length. However, discrepancies between pipeline parameters in practice and their blueprint may lead to inaccurate models \cite{cuiDatadrivenComputationNatural2020}. Obtaining precise friction coefficients of pipelines $\lambda$ is also challenging \cite{marfatiaSteadyStateModelling2022}. Hence, employing a data-driven method to identify precise state-space model for the gas pipeline network through measurement operation data is necessary.

To establish the state-space model for the entire gas pipeline network, we discretize the spatial locations in the partial differential equations \eqref{eq-PDE1}-\eqref{eq-PDE2}. As shown in Fig.~\ref{fig-Schematic}, to depict gas states along a pipeline, it is partitioned into equidistant discrete computation nodes. The states at internal computation nodes are unobservable; only the states at terminal nodes can be measured. Thus, we can only leverage the measurement data at terminal nodes to identify the state-space model of the entire pipeline network. Additionally, the dynamic process of gas cannot be expressed merely as a state transition function due to topological constraints inherent in the gas network, which must be maintained throughout the dynamic process. These characteristics present significant technical challenges in identifying the state-space model of the gas pipeline network.

%% -----------------------------
%% State-Space Model Identification of Gas Pipeline Network
%% -----------------------------

\section{State-Space Model Identification of Gas Pipeline Network}
\label{sce-Methodology}

\subsection{Mathematical Model of Gas Pipeline Network}

To derive a more practical state-space model of a gas pipeline network, we simplify and discretize the fluid partial differential equations \eqref{eq-PDE1}-\eqref{eq-PDE2} and supplement the topology constraints of the whole network.

The convective term in \eqref{eq-PDE2} can be approximately considered as 0 \cite{zhouEquivalentModelGas2017,kiuchiImplicitMethodTransient1994} and the inclination angle $\alpha$ can also be set to zero, that is

\begin{equation}
	\frac{\partial \rho v^2}{\partial x} \approx 0
	\label{eq-Taylor}
\end{equation}

\begin{equation}
	\sin \alpha=0
	\label{eq-Inclination}
\end{equation}

The Taylor expansion method is adopted \cite{qiDecentralizedPrivacyPreservingOperation2020} for the quadratic term of speed in \eqref{eq-PDE2}, that is:

\begin{equation}
	v^2 \approx 2 v_b v-v_b^2
	\label{eq-Taylor2}
\end{equation}

\noindent where $v_b$ denotes the base value of gas velocity.

Besides, the physical variables gas pressure $\pi$ and mass flow $f$ are usually used to represent the state of gas pipeline.

\begin{equation}
	\pi \approx u^2 \rho
	\label{eq-Pressure}
\end{equation}

\begin{equation}
	f=\rho v A
	\label{eq-MassFlow}
\end{equation}

\noindent where $u$ denotes the speed of sound; $A$ denotes the cross-section area of pipeline. Equation \eqref{eq-Pressure} holds under an approximately adiabatic condition.

Finally, based on the approximation in \eqref{eq-Taylor}-\eqref{eq-Taylor2} and the replacement of variables in \eqref{eq-Pressure}-\eqref{eq-MassFlow}, the partial differential equations \eqref{eq-PDE1}-\eqref{eq-PDE2} are transformed into 

\begin{equation}
	\frac{\partial \pi}{\partial t}=-\frac{u^2}{A} \frac{\partial f}{\partial x}
	\label{eq-PDE3}
\end{equation}

\begin{equation}
	\frac{\partial f}{\partial t}=-A \frac{\partial \pi}{\partial x}-\frac{\lambda v_b}{D} f+\frac{\lambda A v_b^2}{2 u^2 D} \pi
	\label{eq-PDE4}
\end{equation}

Then, the partial differential equations can be discretized in the implicit upwind scheme. For any $l \in \mathcal{L}$ and $i \in \mathcal{S}_l$

\begin{equation}
	\pi_{i, t}+\frac{u^2 \Delta t}{A_l \Delta x_l} f_{i, t}-\frac{u^2 \Delta t}{A_l \Delta x_l} f_{i-1, t}=\pi_{i, t-1}
	\label{eq-DDE1}
\end{equation}

\begin{equation}
	\begin{aligned}
	& \left(1	+\frac{\lambda_l v_{b, l} \Delta t}{D_l}\right) f_{i, t} - \frac{A_l \Delta t}{\Delta x_l} \pi_{i-1, t} \\
	& + \left(
		\frac{A_l \Delta t}{\Delta x_l} - \frac{\lambda_l A_l v_{b, l}^2 \Delta t}{2 u^2 D_l}
	\right) \pi_{i, t}
	= f_{i, t-1}
	\end{aligned}
	\label{eq-DDE2}
\end{equation}

\noindent where $\mathcal{L}$ and $\mathcal{S}$ denote the index set of pipelines and computation nodes, respectively; $\mathcal{S}_l$ denotes the index set of computation node in the pipeline $l$. For the convenience of numerical computation, each pipeline is divided into sections by several computation nodes. 

There are also topology constraints in the gas network. First, the gas pressures of terminal segments are equal to the pressure of terminal node. Denote $\mathcal{N}$ as the index set of terminal nodes. For any $n \in \mathcal{N}$

\begin{equation}
	\pi_{i, t}=\pi_{n, t}^{\mathrm{node}}, 
	i \in \mathcal{S}_n
	\label{eq-PressureNode}
\end{equation}

\noindent where $\pi_{n, t}^{\mathrm{node}}$ denotes the gas pressure of terminal node $n$ at time $t$; $\mathcal{S}_n$ denotes the index set of computation nodes coincident with the terminal node $n$. $\pi_{i, t}$ denotes the gas pressure of computation node $i$ at time $t$.

Then, the sum of mass flow into the node equals the sum of mass flow out of the terminal node. For any $n \in \mathcal{N}$

\begin{equation}
	\sum_{i \in \mathcal{S}_n^{+}} f_{i, t}+f_{n, t}^{\mathrm{inj}}=\sum_{j \in \mathcal{S}_n^{-}} f_{j, t}
	\label{eq-MassFlowNode}
\end{equation}

\noindent where $\mathcal{S}_n^{+}$ and $\mathcal{S}_n^{-}$  denote index set of computation nodes that flow into and out of the terminal node, respectively; $f_{n, t}^{\mathrm{inj}}$ denotes the mass flow injection into node $n$ at time $t$; $f_{i, t}$ denotes the mass flow of computation node $i$ at time $t$.

The compressors are used to boost the gas pressure. For any $k \in \mathcal{D}^\mathrm{comp}$

\begin{equation}
	\pi_{j, t}=\pi_{i, t}+\Delta \pi_{k, t}, 
	i \in \mathcal{S}_k^{+}, 
	j \in \mathcal{S}_k^{-}
	\label{eq-Compressor}
\end{equation}

\noindent where $\mathcal{D}^\mathrm{comp}$ denotes the index set of compressors; $\mathcal{S}_k^{+}$ and $\mathcal{S}_k^{-}$ denote the index of computation node flowing into and out the compressor $k$, respectively; $\Delta \pi_{k, t}$ denotes the boosted gas pressure provided by compressor $k$ at time $t$.

\subsection{State-Space Model of Gas Pipeline Network}

In the gas pipeline network, gas pressure and mass flow rate are two variables representing the state at each computation node. Besides, the states inside the pipelines are unobservable, while the states at terminal nodes are measurable or controllable \cite{chenEnergyCircuitBasedIntegratedEnergy2022}. For example, the gas pressure at the gas source node is controllable and the injection gas mass flow rate at the gas consumption node is controllable. The other terminal nodes without gas injection are classified as injection gas mass flow rate measurable, since their gas injections are specified as 0. Denote $\mathcal{N}_{\pi}$ and $\mathcal{N}_{f}$ as the index set of terminal nodes whose gas pressure and gas injection mass flow rate are measured, respectively. Then, the variables in the gas system can be classified into three types: the control variables $\bm{u}_t$, measurable variables $\bm{y}_t$ and state variables $\bm{h}_t$, whose definitions are shown as follows:

\begin{equation}
	\bm{u}_t:=\underset{n\in \mathcal{N}}{\operatorname{col}}\left(u_{n,t}\right), 
  	u_{n, t}:=\left\{
		\begin{array}{c}
			\pi_{n, t}^{\mathrm{node}}, n \in \mathcal{N}_\pi \\
			f_{n, t}^{\mathrm{inj }}, n \in \mathcal{N}_f
		\end{array}
	\right.
	\label{eq-ControlVariable}
\end{equation}

\begin{equation}
	\bm{y}_t:=\underset{n\in \mathcal{N}}{\operatorname{col}}\left(y_{n, t}\right), 
	y_{n, t}:=\left\{
		\begin{array}{c}
			\pi_{n, t}^{\mathrm{node}}, n \in \mathcal{N}_f \\
			f_{n, t}^{\mathrm{inj }}, n \in \mathcal{N}_\pi
		\end{array}
	\right.
	\label{eq-MeansureVariable}
\end{equation}

\begin{equation}
	\bm{h}_t:=\left[\bm{\pi}_t^{\top}, \bm{f}_t^{\top}\right]^{\top}, 
	\bm{\pi}_t:=\underset{i \in \mathcal{S}}{\operatorname{col}}\left(\pi_{i, t}\right), 
	\bm{f}_t:=\underset{i \in \mathcal{S}}{\operatorname{col}}\left(f_{i, t}\right)
	\label{eq-StateVector}
\end{equation}

\noindent where the operator $\operatorname{col}(\cdot)$ denotes the column vector operator.

Based on the discretized partial differential equations \eqref{eq-DDE1}-\eqref{eq-DDE2}, the topology constrains \eqref{eq-PressureNode}-\eqref{eq-MassFlowNode} and compressor model \eqref{eq-Compressor}, the state-space model of the whole gas system can be denoted as the compact matrix form:

\begin{equation}
	\bm{K}(\bm{\theta})\left[
	\begin{array}{c}
		\bm{h}_t \\
		\bm{\pi}_t^{\mathrm{node}}
	\end{array}\right]=\left[
	\begin{array}{c}
		\bm{S} \bm{h}_{t-1} \\
		\bm{u}_t
	\end{array}\right]
	\label{eq-StateSpaceModel0}
\end{equation}

\begin{equation}
	\bm{y}_t=\bm{H} \bm{h}_t
	\label{eq-OutputFunc}
\end{equation}

\noindent where $\bm{K}(\bm{\theta})$  denotes a sparse matrix with the unknown parameters $\bm{\theta}$ to be identified; $\bm{h}_t$ denotes the vector collecting all the state variables at time $t$; $\bm{\pi}_t^{\mathrm{node}}$ denotes the vector composed of the gas pressure at each terminal node at time $t$; $\bm{u}_t$ denotes the vector composed of the control variables at each node and the gas pressure increment of compressors at time $t$; $\bm{S}$ is a constant matrix used to rearrange and select the elements in $\bm{h}_{t-1}$. $\bm{H}$ is also a constant matrix used to calculate the measurable output variables based on the state variables according to the topology constraints.  The method to construct $\bm{S}$ and $\bm{H}$ can be clearly seen in the following small case.

\begin{figure}[h] 
	\centering
	\includegraphics[width=0.35\textwidth]{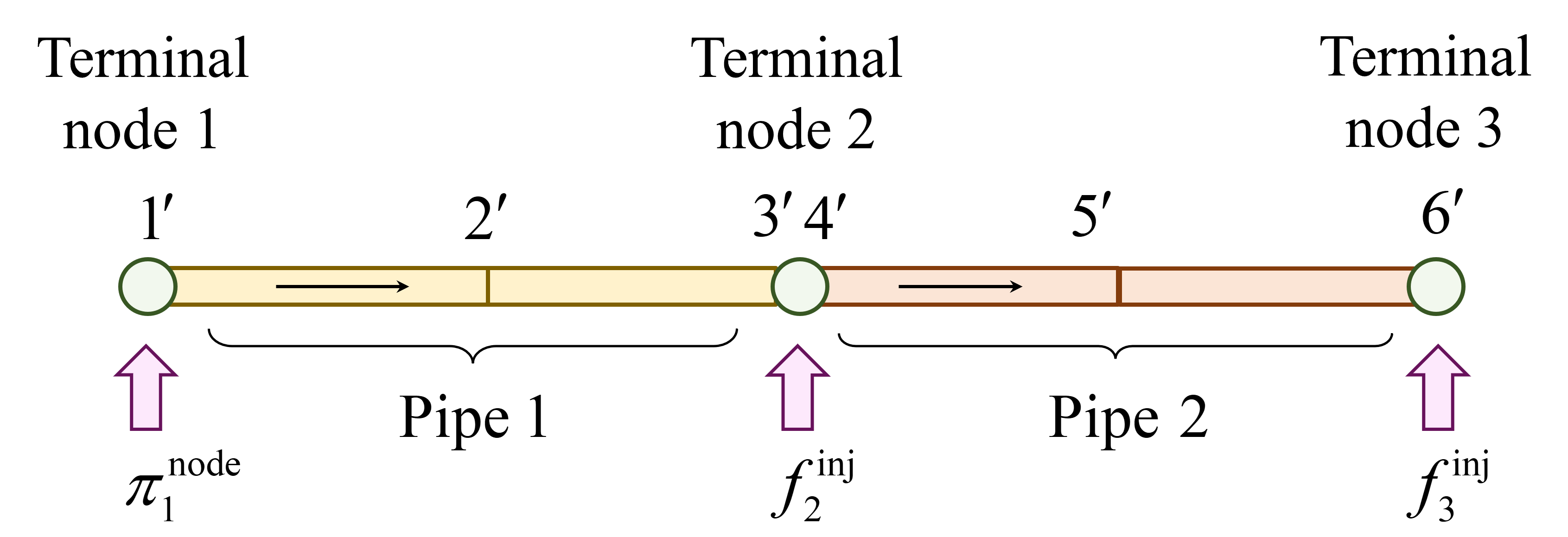}
	\caption{Topology of the 3-node small gas system.} 
	\label{fig-SmallCase}
\end{figure}

To demonstrate the construction of state-space model, we use a 3-node small gas system as an example, as shown in Fig.~\ref{fig-SmallCase}. Each pipeline is divided into 2 segments and there are total of 6 computation points, numbered $1^{\prime}$-$6^{\prime}$ , respectively. 

Suppose the gas pressure at terminal node 1 is controllable, and the injection mass flow rates at terminal node 2 and 3 are controllable. Then, the control variable vector $\bm{u}_t$ is

\begin{equation}
	\bm{u}_t:=\left[
		\pi_{1, t}^{\mathrm{node}}, 
		f_{2, t}^{\mathrm{inj}}, 
		f_{3, t}^{\mathrm{inj}}
	\right]^{\top}
	\label{eq-3NodeControl}
\end{equation}

\begin{figure}[h] 
	\centering
	\includegraphics[width=0.46\textwidth]{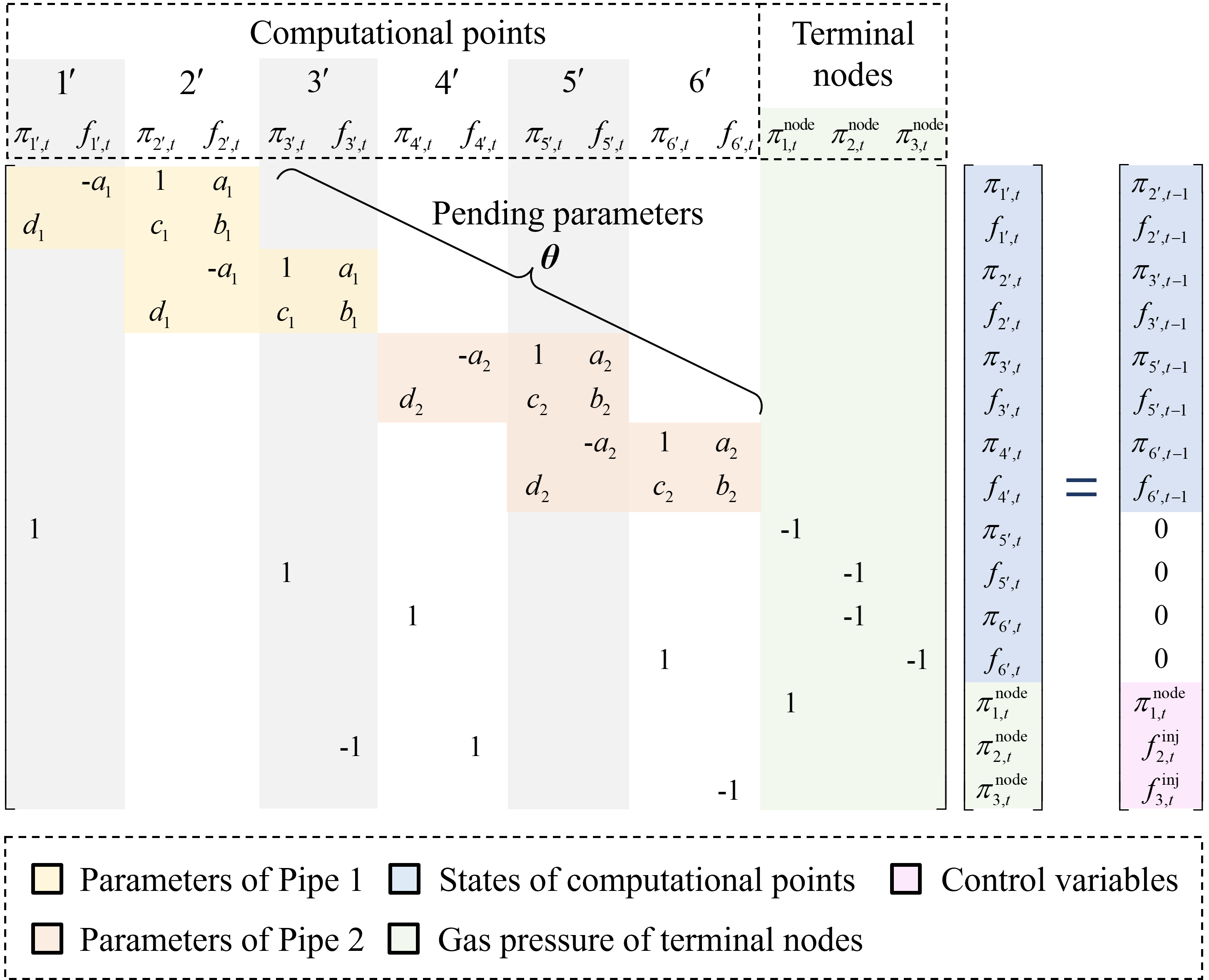}
	\caption{Matrix dynamic function of the 3-node gas system.} 
	\label{fig-MatrixK}
\end{figure}

The dynamic function \eqref{eq-StateSpaceModel0} of this 3-node simple gas system can be written in the following matrix form as shown in Fig.~\ref{fig-MatrixK}.

The detail parameters inside the sparse matrix are defined as follows:

\begin{equation}
	\begin{aligned}
		& a_l:=\frac{u^2 \Delta t}{A_l \Delta x_l}, 
		b_l:=\left(1+\frac{\lambda_l v_{b, l} \Delta t}{D_l}\right) \\
		& c_l:=\left(\frac{A_l \Delta t}{\Delta x_l}-\frac{\lambda_l A_l v_{b, l}^2 \Delta t}{2 u^2 D_l}\right), 
		d_l:=-\frac{A_l \Delta t}{\Delta x_l} \\
		& l \in \mathcal{L}=\{1,2\}
	\end{aligned}
	\label{eq-3NodeParam}
\end{equation}

% 需要注意的是，在参数辨识过程中，$a_l$, $b_l$, $c_l$ and $d_l$ 是作为整体来辨识的，而不是辨识 \eqref{eq-3NodeParam} 中的每个独立参数。

In this case, the parameters $\bm{\theta}$ to be identified, the vector of state variables $\bm{h}_t$, the vector of gas node pressure $\bm{\pi}_t^{\mathrm{node}}$ and the measurable output variable vector $\bm{y}_t$ are shown as follows:

\begin{equation}
	\bm{\theta}:=\underset{l \in \mathcal{L}}{\operatorname{col}}\left(\bm{\theta}_l\right), 
	\bm{\theta}_l:=\left[a_l, b_l, c_l, d_l\right]^{\top}
	\label{eq-3NodeTheta}
\end{equation}

  \begin{equation}
    \bm{h}_t:=\left[
		\begin{array}{l}
			\pi_{1^{\prime}, t}, f_{1^{\prime}, t}, \pi_{2^{\prime}, t}, f_{2^{\prime}, t}, \pi_{3^{\prime}, t}, f_{3^{\prime}, t}, \\
			\pi_{4^{\prime}, t}, f_{4^{\prime}, t}, \pi_{5^{\prime}, t}, f_{5^{\prime}, t}, \pi_{6^{\prime}, t}, f_{6^{\prime}, t}
		\end{array}
	\right]^{\top}
	\label{eq-3NodeState}
\end{equation}

\begin{equation}
	\bm{\pi}_t^{\mathrm{node}}:=\left[
		\pi_{1, t}^{\mathrm{node}}, 
		\pi_{2, t}^{\mathrm{node}}, 
		\pi_{3, t}^{\mathrm{node}}
	\right]^{\top}
	\label{eq-3NodePressure}
\end{equation}

\begin{equation}
	\bm{y}_t:=\left[
		f_{1, t}^{\mathrm{inj}}, 
		\pi_{2, t}^{\mathrm{node}}, 
		\pi_{3, t}^{\mathrm{node}}
	\right]^{\top}
	\label{eq-3NodeOutput}
\end{equation}
 
\begin{figure}[h] 
	\centering
	\includegraphics[width=0.4\textwidth]{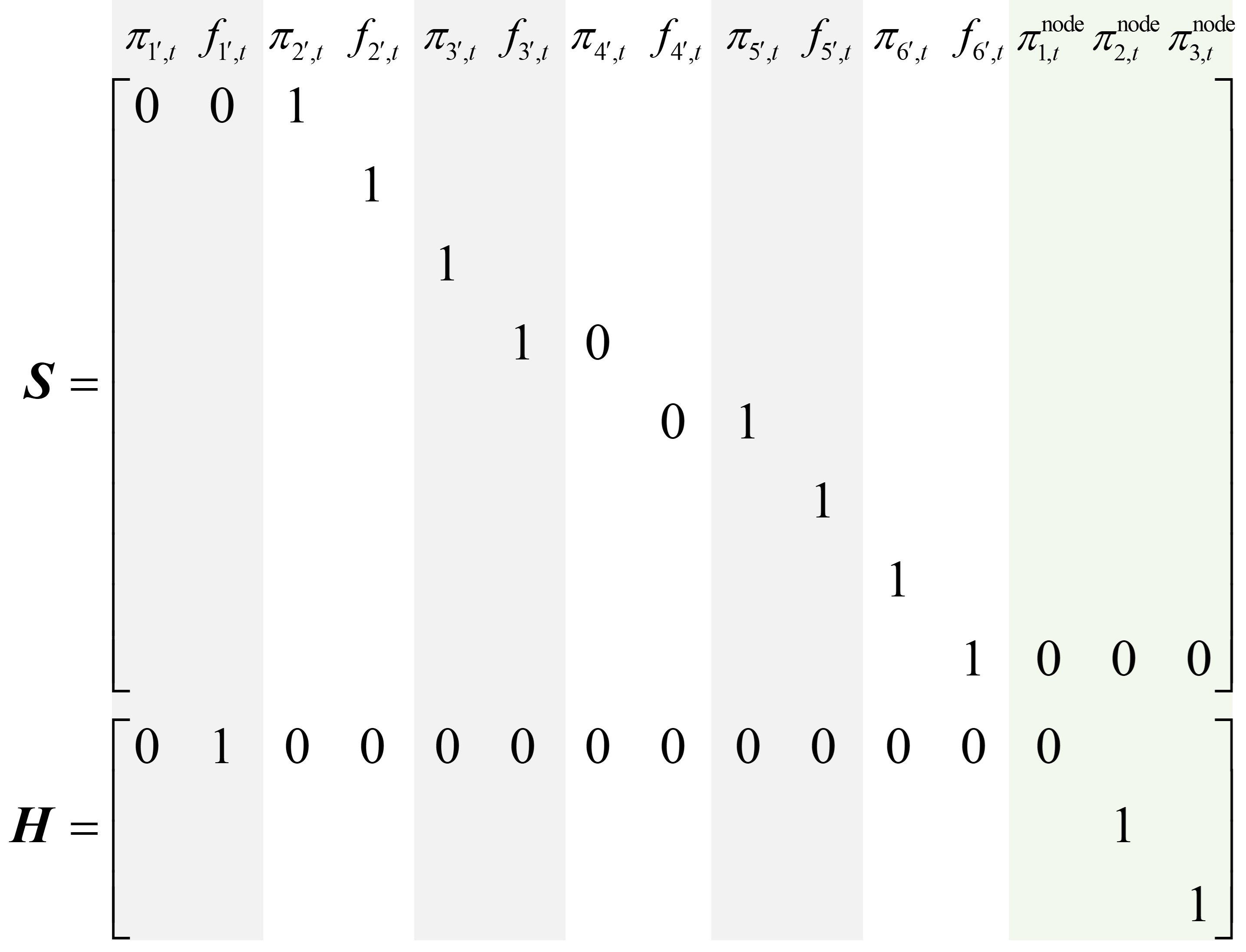}
	\caption{Values of matrices in the 3-node gas system.} 
	\label{fig-MatrixSH}
\end{figure}

According to the matrix dynamic function in Fig.~\ref{fig-MatrixSH}, to calculate the state variables $\bm{h}_t$, it only needs the values in $\bm{h}_{t-1}$ that start from the second computation points. This is because the state of the first computation points of each pipeline ($1^{\prime}$ and $4^{\prime}$ in this case) are not independent. Their values are also determined by topology constraints at terminal nodes. The constant matrix $\bm{S}$ is used to select the independent states from all state variables. In this small case, the value of matrices $\bm{S}$ and $\bm{H}$ are shown in Fig.~\ref{fig-MatrixSH}.

\subsection{Identify the State-Space Model with Physics-Informed Recurrent Network}

This subsection introduces the state-space model of the pipelines using the measurement data in the dynamic process. 
A PIRN is proposed to identify the parameters in the state-space model \eqref{eq-StateSpaceModel0}-\eqref{eq-OutputFunc}. 

The sparse matrix $\bm{K}(\bm{\theta})$ is composed of the dynamic equations \eqref{eq-DDE1} and \eqref{eq-DDE2}, along with the network topology constraints. As a result, it forms a full-rank, invertible matrix. Let the inverse of matrix $\bm{K}(\bm{\theta})$ be denoted as $\bm{J}$, and divide it into four blocks. Then, equation \eqref{eq-StateSpaceModel0} can be transformed into:

\begin{equation}
	\left[\begin{array}{c}
		\bm{h}_t \\
		\bm{\pi}_t^{\mathrm{node}}
	\end{array}\right]=
	\left[\begin{array}{ll}
		\bm{J}^{(1)} & \bm{J}^{(2)} \\
		\bm{J}^{(3)} & \bm{J}^{(4)}
	\end{array}\right]
	\left[\begin{array}{c}
		\bm{S} \bm{h}_{t-1} \\
		\bm{u}_t
	\end{array}\right]
	\label{eq-StateSpace}
\end{equation}

\begin{figure}[h] 
	\centering
	\includegraphics[width=0.46\textwidth]{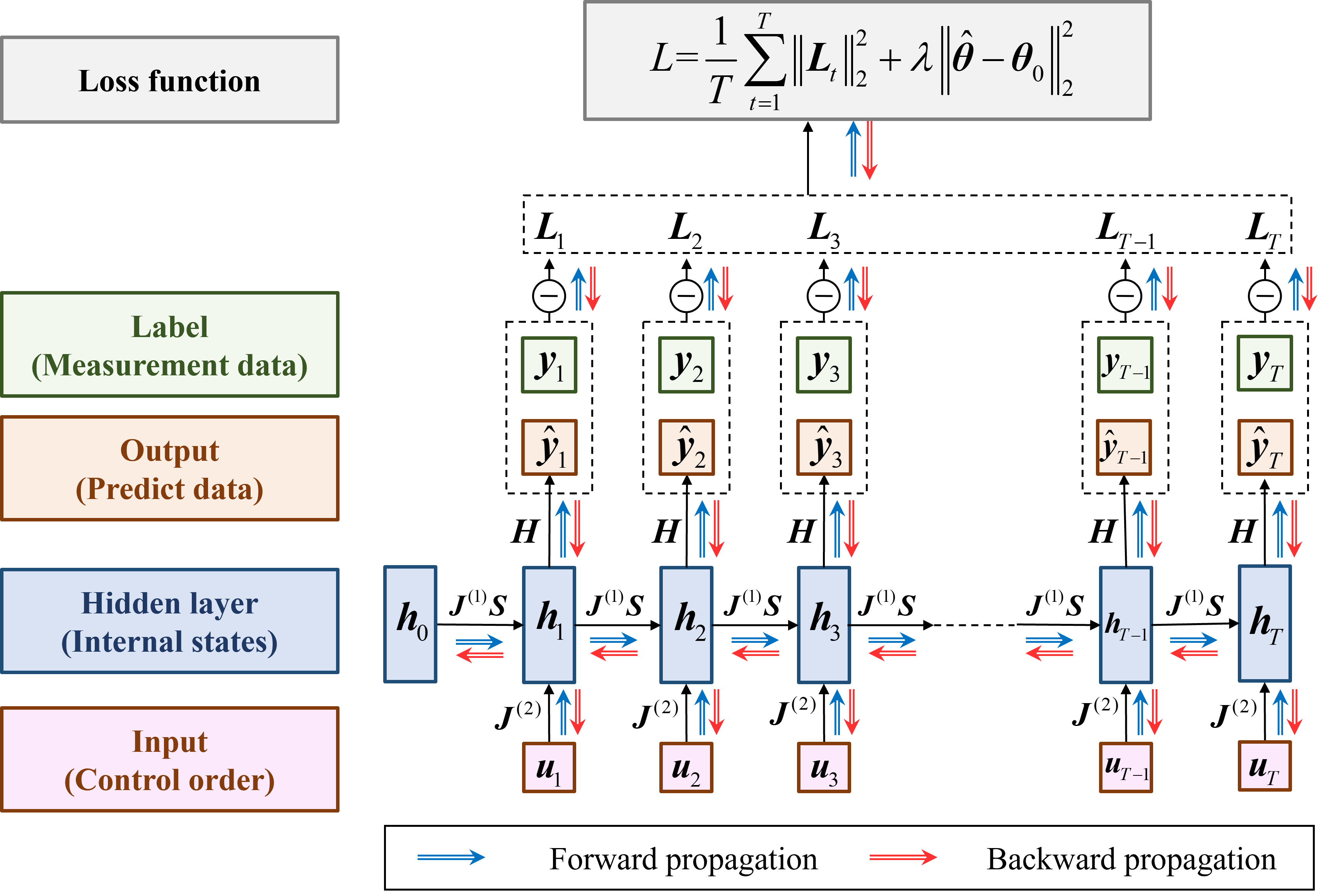}
	\caption{Diagram of PIRN framework.} 
	\label{fig-PIRN}
\end{figure}

Based on the state transition equation \eqref{eq-StateSpace} and the output equation \eqref{eq-OutputFunc}, the diagram of PIRN model training architecture is shown in Fig.~\ref{fig-PIRN}. The control variables $\bm{u}_t$ are inputs of this system. The state variables $\bm{h}_t$ updates based on the input variables and the previous state $\bm{h}_{t-1}$. The outputs $\hat{\bm{y}}_t$ are the predicted output data and the labels $\bm{y}_t$ are the real measured states.  

The loss function is 

\begin{equation}
	L=\frac{1}{T} \sum_{t=1}^T\left\|\bm{L}_t\right\|_2^2+\lambda\left\|\hat{\bm{\theta}}-\bm{\theta}_0\right\|_2^2
	\label{eq-LossFunction}
\end{equation}

\begin{equation}
	\bm{L}_t=\hat{\bm{y}}_t-\bm{y}_t
	\label{eq-LossItem}
\end{equation}

\noindent where $\hat{\bm{\theta}}$ denotes the pending parameters; $\bm{\theta}_0$ denotes the initial estimated values of parameters based on the prior knowledge of physical meaning. The second term of the loss function \eqref{eq-LossFunction} is a regularization term of parameters, and $\lambda$ is a hyper-parameter. Usually, the final solutions obtained by the data-driven learning are not unique, because it may converge to different local optima in the gradient descent process. The regularization term in the loss function can make our training results converge to the parameters that have right physical meaning, rather than other local optima. When the order of magnitude of the estimated parameters $\bm{\theta}_0$ are the same as the real value, the correctness of the parameter solution can be guaranteed.

The goal of backpropagation is to determine the partial derivatives of the loss function with respect to the parameters and initial states \cite{jaegerTutorialTrainingRecurrent2002}, in order to minimize the error between the predicted output data and the measurement data. This allows for the subsequent update of the parameter and initial state values as follows:

\begin{equation}
	\hat{\bm{\theta}} \leftarrow \hat{\bm{\theta}}-\alpha_\theta \frac{\partial L}{\partial \hat{\bm{\theta}}}
	\label{eq-ThetaUpdate}
\end{equation}

\begin{equation}
	\bm{h}_0 \leftarrow \bm{h}_0-\alpha_h \frac{\partial L}{\partial \bm{h}_0}
	\label{eq-InitialUpdate}
\end{equation}  

\noindent where $\alpha_\theta$ and $\alpha_h$ denote the learning rates of $\hat{\bm{\theta}}$ and $\bm{h}_0$, respectively, and they are the hyper-parameters in the model. With the PyTorch framework, we can leverage automatic differentiation to compute partial derivatives numerically, facilitating backward propagation and parameter updates in \eqref{eq-ThetaUpdate}-\eqref{eq-InitialUpdate}. However, this process involves repeated inversion of the large-scale sparse matrix $\bm{K}(\hat{\bm{\theta}})$ to obtain updated $\bm{J}$ as the parameters $\hat{\bm{\theta}}$ updated, which introduces a significant computational burden and can lead to numerical instability.
To address this issue, we apply the matrix inversion lemma to update the new value of the matrix $\bm{J}$. When the parameters $\hat{\bm{\theta}}$ inside the sparse matrix $\bm{K}(\hat{\bm{\theta}})$ change to $\hat{\bm{\theta}}+\Delta\hat{\bm{\theta}}$, the matrix update can be denoted as:

\begin{equation}
	\Delta \bm{K}:=\bm{K}(\hat{\bm{\theta}}+\Delta \hat{\bm{\theta}})-\bm{K}(\hat{\bm{\theta}})=\bm{M} \Delta \bm{\Theta} \bm{N}^{\top}
	\label{eq-MatKUpdate}
\end{equation}

\noindent where $\Delta\bm{\Theta}$ is a diagonal matrix. All parameters change $\Delta \hat{\bm{\theta}}$ within $\bm{K}(\hat{\bm{\theta}}+\Delta \hat{\bm{\theta}})$  are rearranged and assigned to the diagonal elements of the matrix $\Delta \bm{\Theta}$. Constant matrices $\bm{M}$ and $\bm{N}$ are employed to position the parameter changes $\Delta \hat{\bm{\theta}}$ from $\Delta\bm{\Theta}$ to $\Delta \bm{K}$ through matrix operations. Consequently, using the matrix inversion lemma, the updated matrix $J$ can be expressed as

\begin{equation}
	\bm{J} \leftarrow(\bm{K}+\Delta \bm{K})^{-1}=\bm{J}-\bm{J} \bm{M} \bm{\Psi} \bm{N}^{\top} \bm{J}
	\label{eq-MatJUpdate}
\end{equation}

\begin{equation}
	\bm{\Psi}:=\left(\Delta \bm{\Theta}^{-1}+\bm{N}^{\top} \bm{J} \bm{M}\right)^{-1}
	\label{eq-PsiDefinition}
\end{equation}

With the matrix inversion lemma, repeat calculation of large-scale sparse matrix inversion can be effectively avoided. Additionally, it offers greater numerical stability compared to directly computing the backpropagation of an inverse matrix.

In summary, the proposed PIRN model exploits the memory-retention capabilities of RNNs to recover the state-space description of a gas pipeline network from sparse terminal-node measurements. 
Embedding the physical state-space equations within the recurrent architecture ensures interpretability and allows seamless integration into any optimal-dispatch framework.
A comparison between the classic RNNs and the proposed PIRN model is summarized in TABLE \ref{tab-Comparison}.
Besides, to demonstrate the effectiveness of the identified state-space model, the optimal dispatch of the integrated power-gas system is implemented in Section \ref{sec-integrated-dispatch}. 
By jointly optimizing the operation of generators, GTs, GWs, P2Gs and compressors, the economic benefits of the integrated system can be fully exploited.

\begin{table}[h]
	\centering
	\footnotesize
	\caption{Comparison between Classic RNNs and Physics-Informed Recurrent Network}
	\vspace{1em}
	\renewcommand{\arraystretch}{1.5}
	\begin{tabularx}{0.5\textwidth}{>{\centering\arraybackslash}m{0.12\textwidth}>{\centering\arraybackslash}m{0.12\textwidth}>{\centering\arraybackslash}m{0.2\textwidth}}
		\hline
		\hline
		Comparison items & Classic RNNs & Physics-informed recurrent network\\
		\hline
		Input layer & Input data & Control variables order of control units in the gas system\\
		Hidden layer & Hidden states & Internal states of unobservable computation nodes inside the pipelines\\
		Output layer & Predicted data & Predict measurable states of gas\\
		Label & Ground truth & Measurement data of measurable variables\\
		Composition of network & Artificial neural perceptron & Matrices in the state-space function of gas pipeline network\\
		Trainable parameters & Dense matrix without physical meaning & Sparse matrix with specific physical meaning\\
		Training purpose & Classification or fitting & Acquire the parameters of the state-space function\\
		Training algorithm & \multicolumn{2}{c}{Backward propagation through time (BPTT)}\\
		Implementation framework & \multicolumn{2}{c}{PyTorch/TensorFlow} \\
		\hline
		\hline
	\end{tabularx} 
	\label{tab-Comparison}
\end{table}

\subsection{Training Techniques of Physics-Informed Recurrent Network}

Unlike traditional neural network based AI models, the parameters in the PIRN model carry precise physical meaning. 
To ensure both stability and efficiency during training, several specialized techniques must be employed, including warm start, data normalization, gradient clipping and truncated backward propagation through time. 

\subsubsection{Warm Start}

A warm start avoids numerical instability during matrix inversion more efficiently than random initialization, and accelerates the training process. Since all the parameters have an explicit physical meaning, the initial values of pending parameters $\hat{\bm{\theta}}$ and initial states $\bm{h}_0$ can be estimated based on the prior knowledge of physical meanings. The initial value of pending parameters $\hat{\bm{\theta}}$ can be directly set as $\bm{\theta}_0$. The initial states $\bm{h}_0$ in steady state can be estimated according to the physical model of gas pipeline. According to the partial differential equations of gas pipelines, the expressions of mass flow and gas pressure distribution along the pipeline in steady state are as follows:

\begin{equation}
	f(x)=f(0), x \in[0, L]
	\label{eq-MassFlowSteady}
\end{equation}

\begin{equation}
	\pi^2(x)=\pi^2(0)-\frac{\lambda u^2 f^2}{A^2 D} x, x \in[0, L]
	\label{eq-PressureSteady}
\end{equation}

\noindent where $f(x)$ and $\pi(x)$ denote the mass flow and gas pressure at the position $x$ along the pipeline, respectively; $L$ is the total length of the pipeline. Therefore, the initial values of gas pressure $\pi^2(x)$ can be estimated by linear interpolation between the gas pressure at the beginning and end of the pipeline $\pi^2(0)$ and $\pi^2(L)$.

\subsubsection{Data Normalization}

The substantial disparities in pressure and mass flow rate within the gas network, e.g., $10^5$-$10^6$ and $10^0$-$10^1$, pose challenges for training PIRN model. By employing data normalization techniques, diverse physical quantities can be normalized to comparable magnitudes, ensuring numerical stability throughout the training process.

\subsubsection{Gradient Clipping and Truncated Backward Propagation Through Time Method}

Similar to the training process of the classic RNN, the backward propagation through time process also involves a chain of matrix products. High matrix powers may lead to numerical instability, causing gradients to either explode or vanish.
% \cite{zhangDiveDeepLearning2024}. 
The gradient clipping method can be used to solve this problem. In each time step, the gradient is checked and forced to the upper bound or lower bound if it exceeds the limits. Besides, the truncated backward propagation through time method \cite{jaegerTutorialTrainingRecurrent2002} can be applied. This method segments long-term consequences into several shorter ones, reducing the highest matrix powers, thereby enhancing model simplicity and stability.

%% -----------------------------
%% Numerical Tests
%% -----------------------------

\section{Numerical Tests}
\label{sec-NumericalTests}

\subsection{Simulation Setup}

The case studies are carried out on four gas-network topologies of varying scale from the GasLib dataset \cite{schmidtGasLibLibraryGas2017}, including GasLib-11, GasLib-24, GasLib-40, and GasLib-135 topologies. 

The programs are implemented on the Python 3.11 platform using the PyTorch framework.
% \cite{paszkePyTorchImperativeStyle2019}. 
The RMSprop optimizer is employed, with a learning rate ranging from $10^{-5}$ to $10^{-3}$, depending on the specific test cases and quality of  training data. A learning rate scheduler is applied with a step size of 4, while the decay factor (gamma) is set to 0.2. The batch size is configured to 16. All models are trained on a Tesla P100 GPU with 16 GB of VRAM.

In each test case, we first set a group of parameters as the “real” parameters of pipeline network. The finite difference method (FDM) is used to simulate the operation of the gas system. The values of control variables $\bm{u}_t$ are then changed to simulate the dynamic process of gas in the pipeline network and the operation data of all the terminal nodes are collected as the measurement data $\bm{y}_t$.

\subsection{Performance of Pipelines Parameters Identification}

\begin{figure}[h] 
	\centering
	\includegraphics[width=0.45\textwidth]{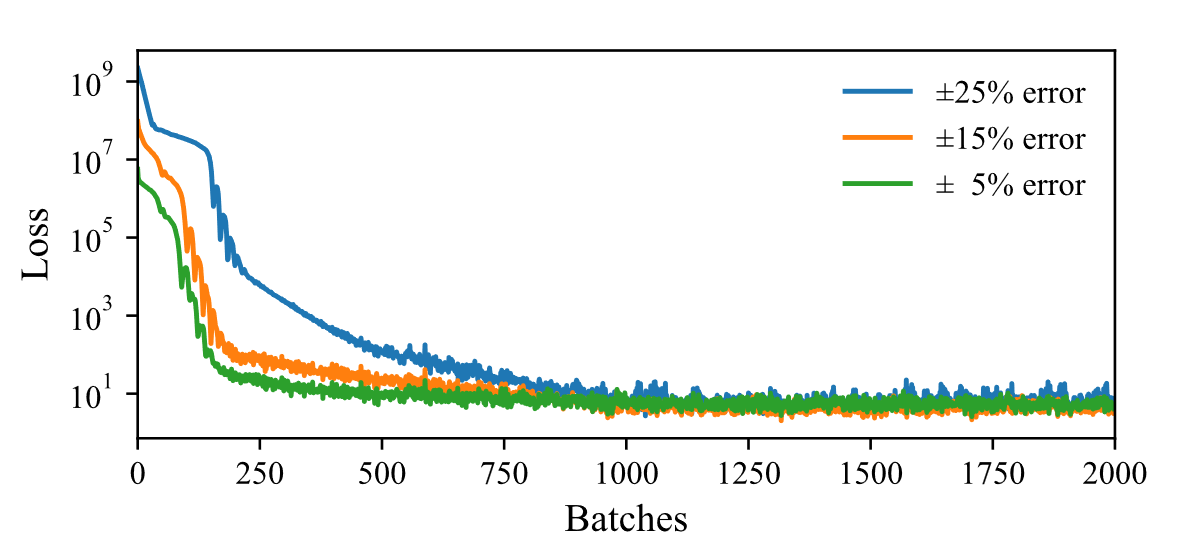}
	\caption{Parameter identification training process of PIRN with different initial estimation errors.} 
	\label{fig-LossInitial}
\end{figure}

To evaluate the performance of the proposed PIRN model, the GasLib-24 system is used as a test case for identifying pipeline parameters. During the warm-start process, the pipeline parameters requiring estimation are initialized with values that contain estimation errors. In this numerical test, three levels of estimation error, within the ranges of ±5\%, ±15\%, and ±25\%, are introduced to assess the model's performance. Random errors are added to the initial estimates within these ranges. The model is then trained for 20 epochs, using measurement data from 100 batches. At the start of each epoch, the data sequence is reshuffled to enhance learning. Fig.~\ref{fig-LossInitial} displays the identification process on a logarithmic scale for different initial estimation errors.
 
Based on the results presented in Fig.~\ref{fig-LossInitial}, the parameter identification process successfully converges across different ranges of initial estimation errors. To assess the performance of the identified state-space model for the gas pipeline network, we employed our model to compute the gas states at each node. These calculated results were then compared to the FDM simulation results. The accuracy was quantified using the mean average percentage error (MAPE) indicator, as presented in TABLE \ref{tab-IdentificationRes}. The results demonstrate that the PIRN model effectively identifies the pipeline parameters, despite varying levels of initial estimation errors.

\begin{table}[h]
	\centering
	\footnotesize
	\caption{Parameters Identification Results}
	\vspace{1em}
	\renewcommand{\arraystretch}{1.5}
	\begin{tabularx}{0.48\textwidth}{>{\centering\arraybackslash}m{0.1\textwidth}>{\centering\arraybackslash}m{0.1\textwidth}>{\centering\arraybackslash}m{0.1\textwidth}>{\centering\arraybackslash}m{0.1\textwidth}}
		\hline
		\hline
		\multirow{2}{*}{MAPE}	&	\multicolumn{3}{c}{Ranges of initial estimation errors}\\
		& ±5\%	&	±15\%	&	±25\% \\
		\hline
		GasLib-11	&	0.13\%	&	0.15\%	& 	0.16\% \\
		GasLib-24	&	0.12\%	&	0.14\%	&	0.19\% \\
		GasLib-40	&	0.11\%	&	0.19\%	&	0.21\% \\
		GasLib-135	&	0.09\%	&	0.12\%	&	0.19\% \\
		\hline
		\hline
	\end{tabularx} 
	\label{tab-IdentificationRes}
\end{table}

\subsection{Robustness under Measurement Data with Outliers}

\begin{figure}[h] 
	\centering
	\includegraphics[width=0.45\textwidth]{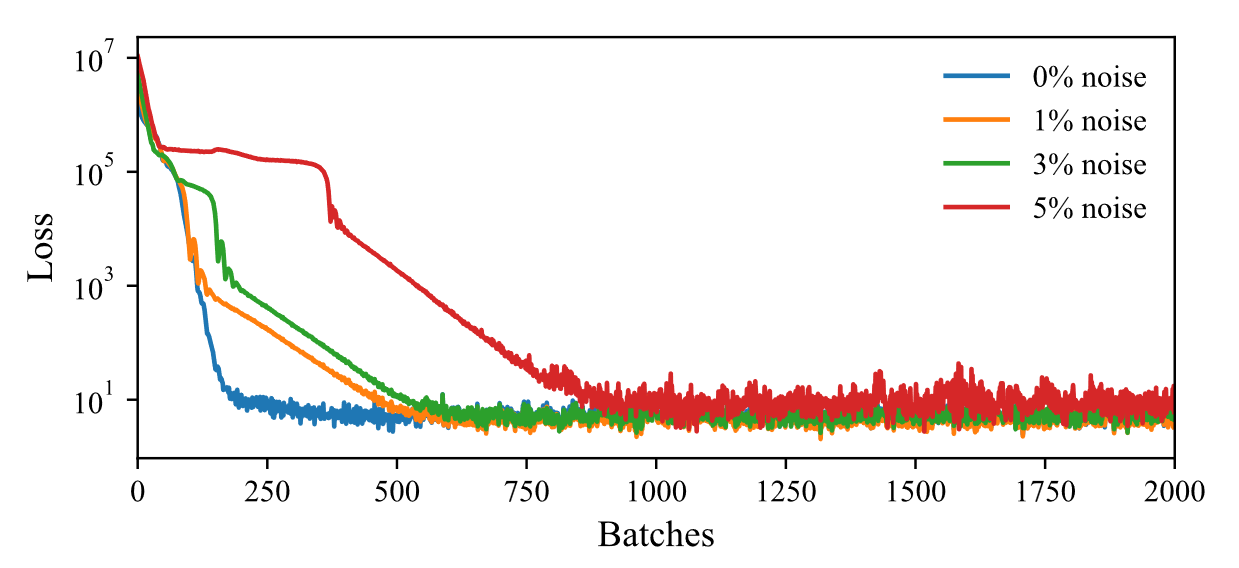}
	\caption{Parameter identification training process of PIRN under different proportions of measurement outliers.} 
	\label{fig-LossNoise}
\end{figure}

In the PIRN model, we employ regularization, gradient clipping, mini-batch stochastic gradient descent, and learning rate scheduling techniques to mitigate the effect of measurement outliers during training. To assess its robustness, outliers with an amplitude of 15\% at varying proportions are randomly introduced to simulate different noise levels in gas network data. Our parameter identification method is then applied to analyze the noise-affected data and identify gas network parameters. Fig.~\ref{fig-LossNoise} illustrates the parameter identification process of PIRN under different noise levels in the GasLib-24 system. As measurement noise deviations increase, the parameters can temporarily be trapped in local optima due to incorrect gradient descent directions affected by measurement outliers, as shown by the 5\% noise curve in Fig.~\ref{fig-LossNoise}. Nevertheless, the model ultimately converges after additional training iterations.

In addition, the MAPE of parameters identification results under different measurement noise with different gas pipeline network systems are listed in TABLE \ref{tab-NoiseRes}.

\begin{table}[h]
	\centering
	\footnotesize
	\caption{Parameters Identification Results under Different Proportions of Outliers}
	\vspace{1em}
	\renewcommand{\arraystretch}{1.5}
	\begin{tabularx}{0.5\textwidth}{>{\centering\arraybackslash}m{0.08\textwidth}>{\centering\arraybackslash}m{0.08\textwidth}>{\centering\arraybackslash}m{0.08\textwidth}>{\centering\arraybackslash}m{0.08\textwidth}>{\centering\arraybackslash}m{0.08\textwidth}}
		\hline
		\hline
		\multirow{2}{*}{MAPE}	&	\multicolumn{4}{c}{Ranges of initial estimation errors}\\
		& 0\%	&	1\%	&	3\%	&	5\%\\
		\hline
		GasLib-11	&	0.11\%	&	0.15\%	&	0.19\%	&	0.31\%\\
		GasLib-24	&	0.15\%	&	0.17\%	&	0.25\%	&	0.37\%\\
		GasLib-40	&	0.13\%	&	0.14\%	&	0.24\%	&	0.33\%\\
		GasLib-135	&	0.12\%	&	0.16\%	&	0.24\%	&	0.35\%\\
		\hline
		\hline
	\end{tabularx} 
	\label{tab-NoiseRes}
\end{table}

\subsection{Parameter Efficiency Compared to Classic Neural Networks}

The proposed PIRN integrates the dynamic state-space function of the gas pipeline network, resulting in a model with an explicit matrix operation form and a relatively small number of required parameters to achieve strong fitting performance. 
In this test case, we compare the PIRN model with the classic Elman RNN and LSTM networks. All networks are trained using control variables $\bm{u}_t$ as inputs and measurement data $\bm{y}_t$ as output labels.

When all models are required to reach fitting errors within the same order of magnitude, we evaluate and compare the number of parameters necessary for each model.

\begin{figure}[h] 
	\centering
	\includegraphics[width=0.45\textwidth]{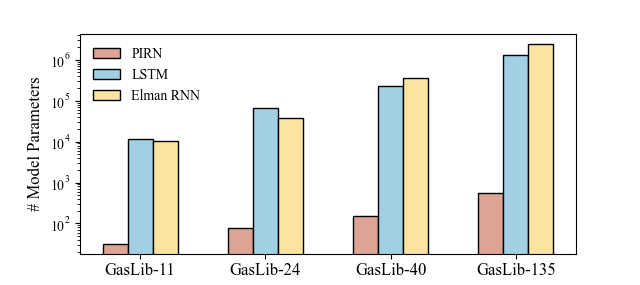}
	\caption{Number of parameters required for each model.} 
	\label{fig-ParamEffc}
\end{figure}

According to Fig.~\ref{fig-ParamEffc}, under the condition of achieving similar fitting accuracy, both Elman RNN and LSTM networks require significantly more parameters compared to the proposed PIRN. Moreover, as the network scale increases, this disparity becomes increasingly pronounced. Therefore, compared to RNN and LSTM networks, the proposed PIRN demonstrates better parameter efficiency compared to the classic neural networks.

\subsection{Integrate the Identified State-Space Model in the Gas-Electricity Integrated Optimal Dispatch Problem}
\label{sec-integrated-dispatch}

\begin{figure}[h] 
	\centering
	\includegraphics[width=0.45\textwidth]{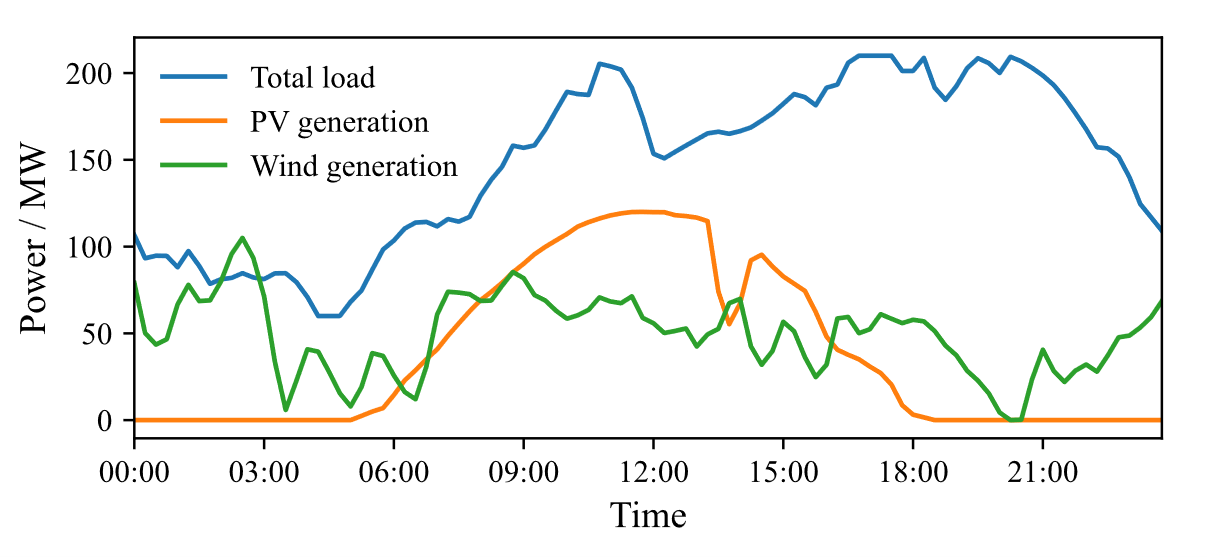}
	\caption{Total load and renewable generation curves.} 
	\label{fig-SetupCurves}
\end{figure}

The PIRN-based parameter identification method produces an explicit state-space function for the gas pipeline network as in \eqref{eq-StateSpace}, which can directly integrate in the gas-electricity integrated optimal dispatch problem. In this subsection, the IEEE RTS96 One Area 24-bus power system and the 24-pipe benchmark gas system are used as numerical test cases to demonstrate how the identification results can be integrated into the optimal dispatch problem. The detailed configuration and topology of this system can be found in \cite{zlotnikCoordinatedSchedulingInterdependent2017}. The gas system includes 4 GTs, 4 P2Gs, and 5 compressors, while the power system consists of 1 coal generator, 3 wind turbines, and 2 photovoltaic units. Since modeling of gas-electricity integrated optimal dispatch is not the main focus of this paper, the detailed descriptions of the system configurations and mathematical models can be found in the supplementary file.
% \cite{siyuanwangSupplementalFilePhysicsInformed}. 
The total power system load and renewable generation curves are presented in Fig.~\ref{fig-SetupCurves}.
 
\begin{figure}[h] 
	\centering
	\includegraphics[width=0.45\textwidth]{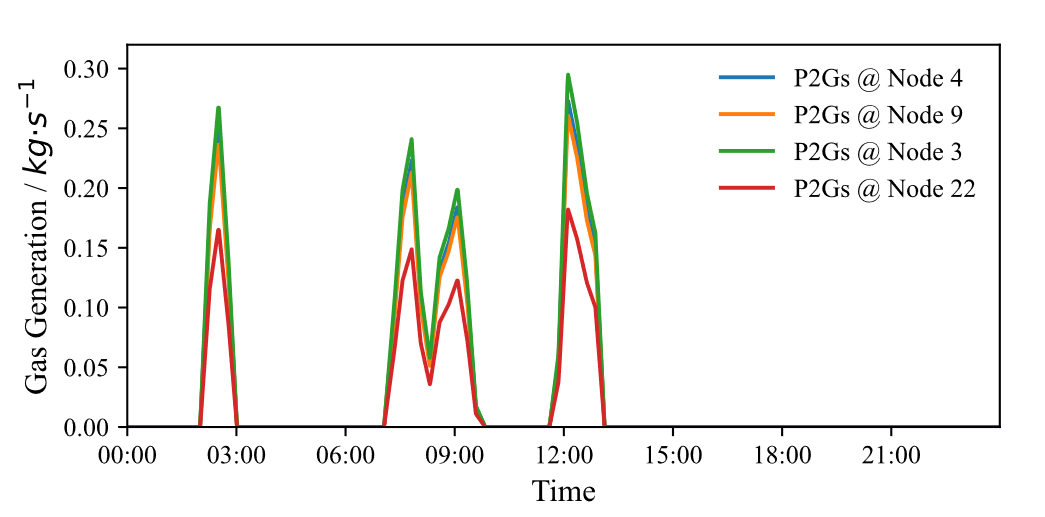}
	\caption{Gas generation of P2Gs throughout the day.} 
	\label{fig-P2Gcurves}
\end{figure}

\begin{figure}[h] 
	\centering
	\includegraphics[width=0.45\textwidth]{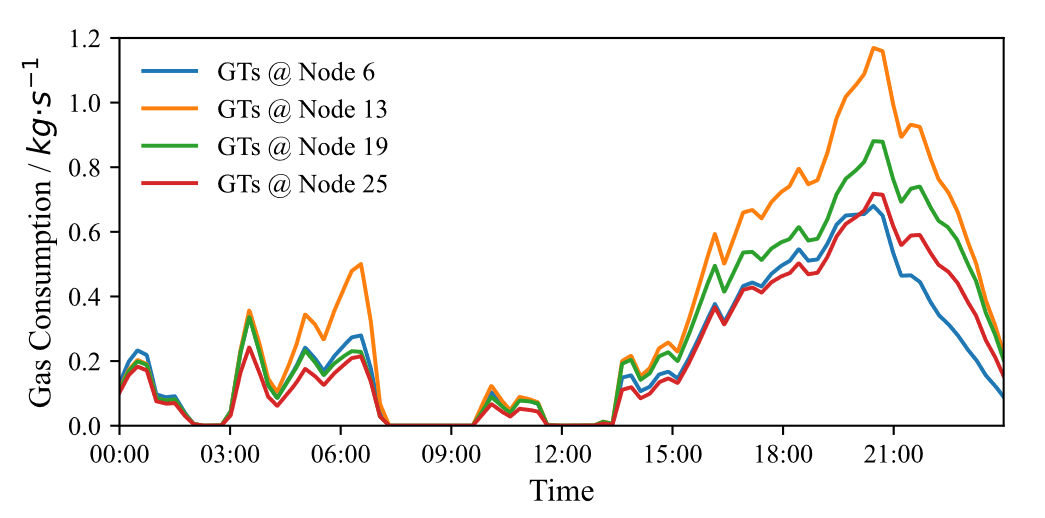}
	\caption{Gas consumption of GTs throughout the day.} 
	\label{fig-GTCurves}
\end{figure}

Fig.~\ref{fig-P2Gcurves} and Fig.~\ref{fig-GTCurves} display the gas generation curves of P2Gs and gas consumption of GTs throughout the day, respectively. During three specific timeframes (2:00-3:00, 7:00-10:00, and 11:30-13:00), the output power from renewable energy generators surpasses the total power load of the system. Consequently, P2Gs undergo electrolysis to produce natural gas for energy storage purposes. During these intervals, GTs stop gas consumption for electricity generation.
 
\begin{figure}[h] 
	\centering
	\includegraphics[width=0.45\textwidth]{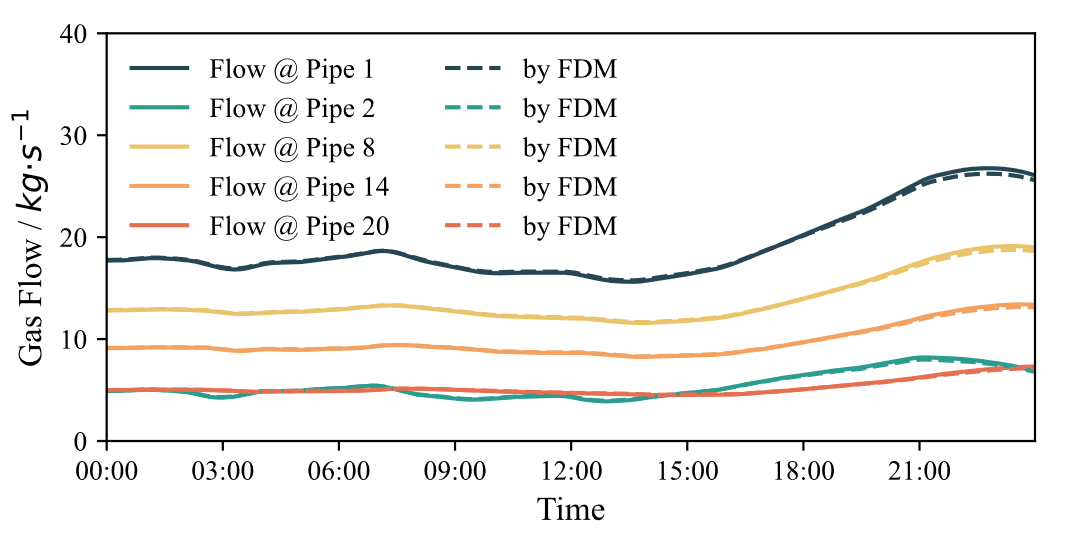}
	\caption{Gas pressures of several terminal nodes throughout the day.} 
	\label{fig-MassFlow}
\end{figure}

\begin{figure}[h] 
	\centering
	\includegraphics[width=0.45\textwidth]{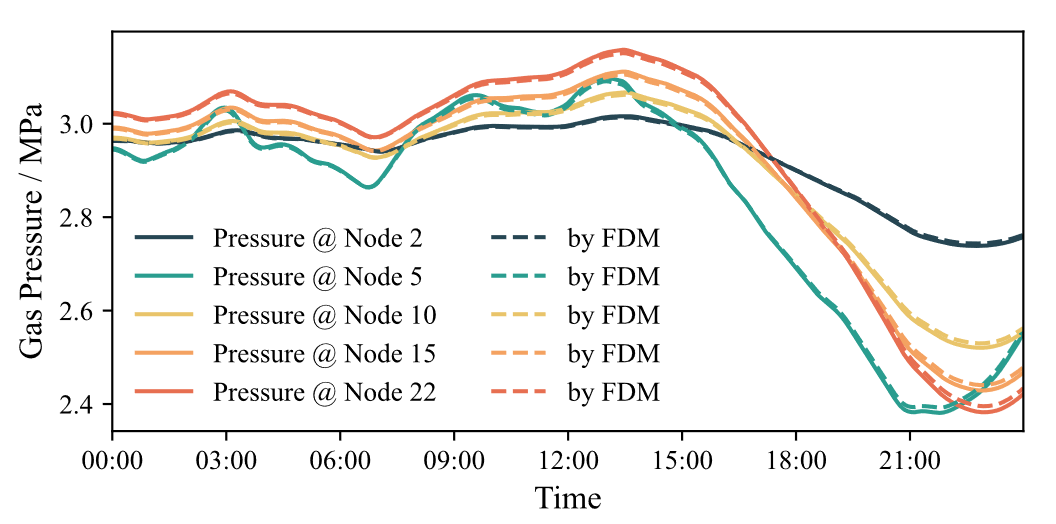}
	\caption{Gas flow rates of several pipelines throughout the day.} 
	\label{fig-GasPress}
\end{figure}

Fig.~\ref{fig-MassFlow} and Fig.~\ref{fig-GasPress} displays the gas mass flow rates in multiple pipes and the gas pressures at various nodes throughout the day, respectively. The solid lines represent our model's calculated results, contrasting with the dashed lines reflecting outcomes from the FDM results. Then, we use the statistical indicators MAPE, RMSE and $R^2$ to demonstrate the performance of our method. Error statistics of gas pressure and gas mass power flow throughout the day are shown in TABLE \ref{tab-DispatchRes}. The identified state-space model can represent the dynamic process of the gas system with high calculation accuracy. Therefore, it can be effectively embedded in the scenarios such as optimal dispatch of integrated energy systems.

\begin{table}[h]
	\centering
	\footnotesize
	\caption{Error Statistics of Gas Pressure and Gas Mass Power Flow}
	\vspace{1em}
	\renewcommand{\arraystretch}{1.5}
	\begin{tabularx}{0.45\textwidth}{>{\centering\arraybackslash}m{0.09\textwidth}>{\centering\arraybackslash}m{0.15\textwidth}>{\centering\arraybackslash}m{0.15\textwidth}}
		\hline
		\hline
		\multirow{2}{*}{Indicators}	&	\multicolumn{2}{c}{Statistics of dispatch results}\\
		& Gas pressure	&	Gas mass flow rate \\
		\hline
			MAPE	&	0.1762\%	&	0.2131\% \\
			RMSE	&	0.0014 MPa	&	0.0374 kg/s \\
			$R^2$	&	0.9999	&	0.9989 \\
		\hline
		\hline
	\end{tabularx} 
	\label{tab-DispatchRes}
\end{table}

%% -----------------------------
%% Conclusion
%% -----------------------------

\section{Conclusion}
\label{sec-Consclusion}

This paper introduces a physics-informed recurrent network model to identify the state-space model of the gas pipeline network using sparse measurement data from terminal nodes of pipelines. It combines data-driven learning from measurements with fluid partial differential equations governing the dynamic states of the gas pipeline network. This PIRN incorporates a physical embedding by substituting artificial neural perceptron units with the gas pipeline network's physical state-space model, ensuring physical interpretability. 

Utilizing the PyTorch framework, the parameters of the state-space model can be accurately learned. Case studies demonstrate that our proposed framework accurately identifies the gas pipeline network model and simultaneously estimates the internal states of gas along the pipelines. Additionally, the identified state-space model can be conveniently applied in the joint optimal dispatch of the integrated gas-electricity system problem and it achieves high accuracy and parameter efficiency.

The PIRN framework uses far fewer parameters than classic neural networks, making it more sensitive to individual parameter values. This heightened sensitivity presents a challenge in quickly determining optimal hyper-parameters to improve convergence speed, requiring further exploration. Additionally, extending the PIRN framework to more complex models of gas pipeline networks, such as nonlinear models, is another important avenue for future research.

\bibliographystyle{IEEEtran}
\bibliography{references}

% \newpage

 

% \end{IEEEbiography}


  
% \vfill

\end{sloppypar}
\end{document}